\documentclass[letter]{aa}  

\usepackage{graphicx}
\graphicspath{{./}{figures/}}
\usepackage{natbib}
\usepackage{amsmath}
\usepackage{mathrsfs}
\usepackage{xcolor}
\usepackage{txfonts}
\usepackage[colorlinks=true]{hyperref}
\hypersetup{linkcolor=blue,citecolor=blue,filecolor=blue,urlcolor=blue}
\usepackage[normalem]{ulem}  
\begin{document}
\newcommand*{\paolam}{\textcolor{magenta}}

\title{Quasar Main Sequence unfolded by 2.5D FRADO}

\subtitle{(Natural expression of Eddington ratio, black hole mass, and inclination)}

   \author{M. H. Naddaf
          \inst{1}\fnmsep\thanks{mh.naddaf@uliege.be}
          \and
          M. L. Mart{\'i}nez-Aldama
          \inst{2,3,4}
          \and
          P. Marziani
          \inst{4}
          \and
          B. Czerny
          \inst{5}
          \and
          D. Hutsem\'ekers
          \inst{1}
}

   \institute{Institut d'Astrophysique et de Géophysique, Université de Liège, Allée du six août 19c, B-4000 Liège (Sart-Tilman), Belgium
         \and
    Astronomy Department, Universidad de Concepción, Barrio Universitario s/n, Concepción 4030000, Chile
        \and
        Millennium Nucleus on Transversal Research and Technology to Explore Supermassive Black Holes (TITANs)
        \and 
        Millennium Institute of Astrophysics (MAS), Nuncio Monseñor Sótero Sanz 100, Providencia, Santiago, Chile
         \and
    National Institute for Astrophysics (INAF), Astronomical Observatory of Padua, Vicolo Osservatorio 5, IT35122, Padua, Italy
         \and
    Center for Theoretical Physics, Polish Academy of Sciences, Al. Lotnik\'ow 32/46, 02-668 Warsaw, Poland
             }

   \date{Submitted: August, 2025}

 
  \abstract
   {}
  {The quasar main sequence (QMS) -- characterized by the Eigenvector 1 (EV1) -- serves as a unifying framework for classifying type-1 active galactic nuclei (AGNs) based on their diverse spectral properties. Although it has long eluded a fully self-consistent physical interpretation, our physically motivated 2.5D FRADO (Failed Radiatively Accelerated Dusty Outflow) model now naturally predicts that Eddington ratio ($\dot{m}$) is the underlying physical primary driver of QMS, with black hole mass ($M_{\bullet}$) and inclination ($i$) acting as secondary contributors.}
   {We recruited a dense grid of FRADO simulations of the geometry and dynamics of the broad-line region (BLR) covering a representative range of $M_{\bullet}$ and $\dot{m}$. For each simulation, we computed the full width at half maximum (FWHM) of the H$\beta$ line under different $i$.}
   {The resulting FWHM–$\dot{m}$ diagram strikingly resembles the characteristic trend observed in the EV1 parameter space. Therefore, it establishes the role of $\dot{m}$ as the true proxy for Fe\textsc{ii} strength parameter ($R_{\rm{Fe}}$), and vice-versa. Our results suggest that $\dot{m}$ can be the sole underlying physical tracer of $R_{\rm{Fe}}$ and should therefore scale directly with it. The $M_{\bullet}$ accounts for the virial mass–related scatter in FWHM. The $i$ then acts as a secondary driver modulating the $R_{\rm{Fe}}$ and FWHM for a given $\dot{m}$ and $M_{\bullet}$, respectively.}
   {}


   \maketitle


\section{Introduction}

AGNs are among the most luminous and energetic astrophysical phenomena, driven by accretion of matter onto supermassive black holes (SMBHs) at the centers of galaxies. The vast diversity in their observed properties -- including radio-loud (RL) and radio-quiet (RQ) classes, broad and narrow emission lines, and differences in continuum shapes -- is believed to result from variations in fundamental physical parameters such as black hole mass ($M_{\bullet}$), Eddington ratio ($\dot{m}$), black hole spin ($a$), and the inclination ($i$) of the accretion disk relative to the observer’s line of sight, measured from the disk symmetry axis \citep[][see Appendix~\ref{sec:accretion_definition} for Notations]{Netzer2015, Padovani2017}.

A major breakthrough in organizing this spectral diversity was made by \citet{boroson1992}, who performed a principal component analysis (PCA) on a sample of low-redshift quasars. Their work identified a dominant eigenvector—later termed EV1 -- that captured strong correlations among several key optical spectral features. Most notably, EV1 revealed an anti-correlation between the optical Fe\textsc{ii} emission and the O\textsc{iii} [$\lambda5007$] line, as well as a correlation between Fe\textsc{ii} strength and the FWHM of the broad H$\beta$ emission line. These connections formed the basis for the QMS, a concept that has since become central to classification of Type-1 AGNs \citep{Sulentic2000, Sulentic2000b, Shen2014, Marziani2018, panda2019b}.

The Fe\textsc{ii} strength parameter, $R_{\rm{Fe}}$, defined as the ratio of the equivalent width (EW) of optical Fe\textsc{ii} blend (4434–4684~\AA) to the EW of broad H$\beta$, is the metric for locating a quasar along the QMS \citep{Sulentic2000, Shen2014, Marziani2018}. When plotted in a plane spanned by FWHM(H$\beta$) and $R_{\rm{Fe}}$, the QMS provides a two-dimensional framework for understanding the spectroscopic diversity of quasars. This diagram allows the classification of quasars into Population A (PA) and Population B (PB) sources. PA ($\mathrm{FWHM}_{\rm H\beta}$ $\lesssim$ 4000 km~s$^{-1}$) objects typically exhibit Lorentzian profiles, strong Fe\textsc{ii} emission, softer X-ray spectra, and are predominantly narrow-line Seyfert 1 galaxies (NLS1s), while PB ($\mathrm{FWHM}_{\rm H\beta}$ $\gtrsim$ 4000 km~s$^{-1}$) sources show Gaussian profiles and tend to have weaker Fe\textsc{ii}, harder X-ray spectra, and contain more frequent radio-loud sources than the other \citep{Sulentic2000b, FraixBurnet2017, Berton2020}.

Over the past three decades, the QMS framework has been validated and extended by large-scale optical and ultraviolet spectroscopic surveys. The relationships identified by \citet{boroson1992} are found to persist across a broad range of redshifts and luminosities, from bright, low-redshift quasars to fainter, distant sources.  These consistent trends suggest the QMS may serve, like the Hertzsprung–Russell diagram for stars, as a tool for tracking the physical and evolutionary states of AGNs.

However, while the QMS is well established empirically, its underlying physical driver has long been debated. Increasing evidence suggests that the Eddington ratio $\dot{m}$ (often expressed in the literature as $\lambda_{\mathrm{Edd}}$; see Appendix~\ref{sec:accretion_definition} for the equivalence) is the primary variable regulating EV1, with $M_{\bullet}$ and $i$ playing secondary roles \citep{Boroson2002, Marziani2018}. This interpretation is further supported by studies that show strong correlations between $R_{\rm{Fe}}$ and X-ray spectral properties, offering a more direct connection to accretion physics \citep{Dupu2019, Panda2018}. In particular, the strength of Fe\textsc{ii} emission has been proposed as a useful surrogate for the accretion rate, providing a practical observational handle on the central engine of Type-1 AGNs.

While the QMS is an effective empirical classification scheme, its structure is rooted in the physics of the BLR, which produces the H$\beta$ and Fe\textsc{ii} emission lines defining its parameters. Understanding the BLR’s formation, structure, and dynamics is key to explaining trends along the QMS and EV1, as variations in $\dot{m}$, and $M_{\bullet}$ shape BLR cloud formation, spatial kinematics. Previous BLR models have been largely parametric or phenomenological, prescribing geometries and velocity fields without linking them to AGN physics. Lacking direct ties to physical parameters like $\dot{m}$ and $M_{\bullet}$, they cannot fully explain EV1 trends. A self-consistent model connecting these physical parameters to the BLR is therefore essential to relate the QMS structure to the central engine’s physics.

The FRADO (Failed Radiatively Accelerated Dusty Outflow) model \citep{Czerny2011} provides a physically motivated mechanism for BLR formation, where radiation pressure on dusty gas lifts material from the dust-rich regions at large radii of accretion disk. As dust grains sublimate, the wind stalls, producing a bound “failed” outflow that forms the BLR. This links BLR structure to radiation field governed by $M_{\bullet}$, and $\dot{m}$. Originally one-dimensional, FRADO has been extended to a 2.5D version \citep{naddaf2021, naddaf2022} that includes vertical and radial cloud dynamics, modeling the effects of gravity, radiation pressure, and dust sublimation on trajectories. This extension predicts BLR geometry, velocity fields, and line profiles directly from $M_{\bullet}$, and $\dot{m}$, etc.

In this study, we show that by letting $\dot{m}$ serve as $R_{\rm{Fe}}$, the 2.5D FRADO model reproduces the observed QMS structure, providing the first physically grounded explanation for quasar distribution along EV1. Section \ref{sec:method} outlines our physically based approach to deriving the QMS structure, Section \ref{sec:results} presents the key results, Section \ref{sec:discussion} offers a detailed discussion, and Section \ref{sec:conclusion} summarizes our conclusions.

\section{Methodology}\label{sec:method}

We utilized a dense grid of FRADO simulations \citep[reported in our very recent work][]{naddaf2025}, which self-consistently model the geometry and dynamics of the BLR. The simulation set spans a representative range of $M_{\bullet}$, $\dot{m}$, and $i$, which is performed for metallicity (Z) of $5 Z_{\odot}$ where $Z_{\odot}$ is the solar value. From this grid of simulations, we extracted the FWHM of the H$\beta$ emission line for each configuration \citep[see][for details]{naddaf2025}. We then analyzed the dependence of $\mathrm{FWHM}_{\rm H\beta}$ on these parameters. This investigation led to a compelling, physically motivated outcome about the QMS, which we present in details in the following sections. It should be noted that results for Fe\textsc{ii} emission are still pending, given the intricacies involved in its modeling, especially when compared to the more tractable approach we adopted for modeling H$\beta$ line \citet{naddaf2025}.

\section{Results}\label{sec:results}

\subsection{FWHM --~$\dot{m}$ plane}
Figure \ref{fig:eigenvector} shows the simulated $\mathrm{FWHM}_{\rm H\beta}$ as a function of $\dot{m}$. The data points are color-coded by (left) $\log M_{\bullet}$ and (right) shape factor ($D_{\rm{H}\beta}$; defined as the ratio of FWHM to the line dispersion, $\sigma$). They are also marked according to three inclination angles of $20^\circ$, $39^\circ$, and $60^\circ$. The simulations overlap remarkably well with the locus occupied by observational points in the QMS. The division between PA and PB is naturally recovered, as is the locus of NLS1 sources. The $\mathrm{FWHM}_{\rm H\beta}$ shows a clear positive correlation with $M_{\bullet}$ and a negative one with $\dot{m}$, as evident in the Figure \ref{fig:eigenvector}.
Interestingly, the simulations also show that PA sources are of more Lorentzian profiles \textbf{($D_{{\rm H}\beta}\rightarrow 0$)}, while PB tend to be described by broader profiles of more Gaussian type \textbf{($D_{{\rm H}\beta} \sim 2.35$)} \citep{zamfir2010}. Moreover, our previous studies \citep{naddaf2022, naddaf2025} show that as $\dot{m}$ rises, more line asymmetry and line shift are expected.

\subsection{Scaling of $R_{\rm{Fe}}$ with $\dot{m}$}\label{sect:rfe_mdot}

\begin{figure*}
        \centering
        \includegraphics[scale=0.45]{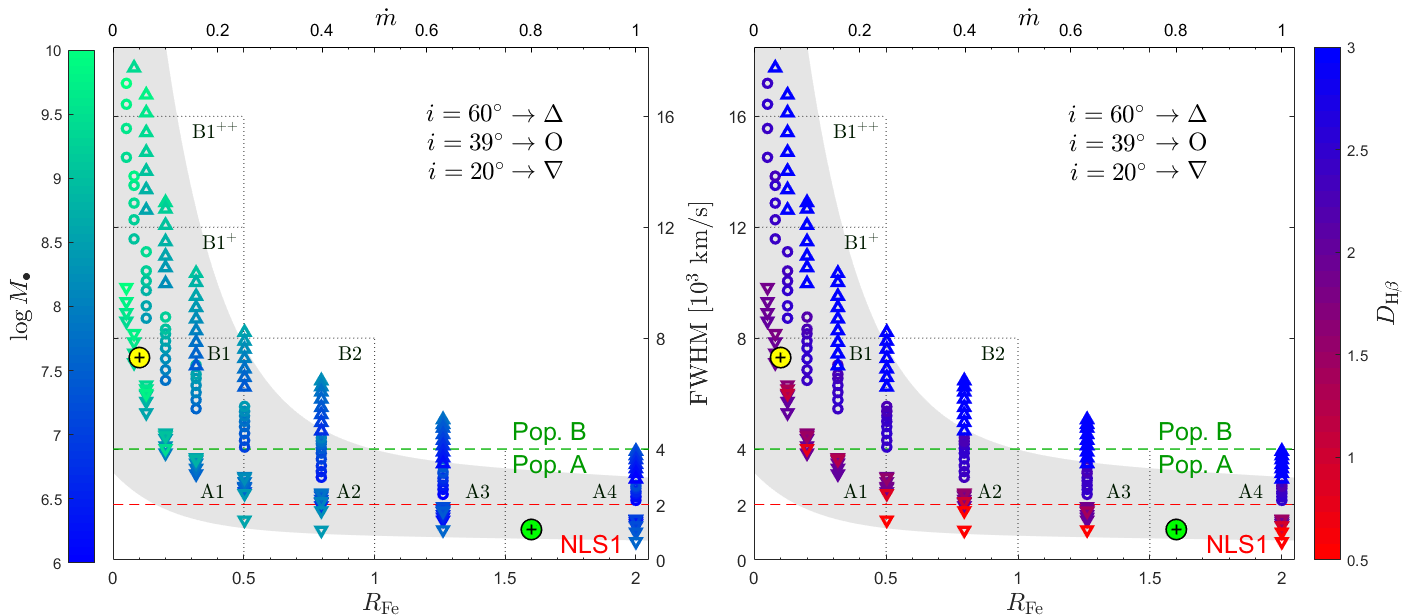}
        \caption{EV1 expressed with $\dot{m}$ as the proxy for $R_{\rm{Fe}}$. Points show $\mathrm{FWHM}_{\rm H\beta}$ vs. $\dot{m}$ from FRADO-driven BLR models at different inclinations, color-coded by (left) $\log M_{\bullet}$ and (right) $D_{\rm{H}\beta}$. The upper and lower horizontal axes correspond to FRADO and observational data (gray-shaded area), respectively, with no scaling relation applied between them. Green and yellow circles ($\oplus$) mark the $R_{\rm{Fe}}$-based locations of I Zw 1 \citep{Marziani2021} and NGC 5548 \citep{Dupu2019}. The horizontal green dashed line marks the PA/PB boundary, and the region below the red dashed line indicates the location of NLS1 objects.}
        \label{fig:eigenvector}
\end{figure*}

Comparing our results in the FWHM–$\dot{m}$ plane with the typical QMS diagram reveals a strong resemblance between FRADO predictions and observed trends \citep[see][and references therein]{zamfir2010}. As shown in Figure~\ref{fig:eigenvector}, this similarity indicates that $\dot{m}$ and $R_{\rm{Fe}}$ directly trace each other, implying a clear physical scaling without requiring additional parameters, not even $Z$ (see Section~\ref{subsec:metallicity}). In other words, higher $\dot{m}$ naturally leads to stronger Fe\textsc{ii} emission relative to H$\beta$, i.e., a larger $R_{\rm{Fe}}$.

However, empirical relations between these quantities — including those from \citet{dupu2016L} and our own fits to the \citet{Hu2008} and \citet{Wu2022} samples (Appendix~\ref{sect:appendixA}) — should be treated with caution. Estimates of $\dot{m}$, $M_{\bullet}$, and bolometric luminosity ($L_{\rm bol}$) are prone to significant biases and depend strongly on the chosen derivation method \citep[see e.g.][]{kaspi2000, onken2004, Hu2008, Jin2012, ho_kim2014, dupu2016L, Chen2022, Wu2022}. Discrepancies arise from differences in bolometric corrections (BC), virial factors, spectral data quality, observational mode (single- vs. multi-epoch), and especially the unknown inclination angle. Thus, while $R_{\rm{Fe}}$ and $\dot{m}$ are positively correlated, empirical calibrations are highly sample-dependent and sensitive to data quality. Appendix~\ref{sect:appendixA} provides a discussion of these limitations and biases.

\section{Discussion}\label{sec:discussion}

The results presented here show that the 2.5D FRADO model is capable of reproducing the main structure of the quasar MS in the optical EV1 plane. The progression from broad H$\beta$, weak Fe\textsc{ii} emitters to narrower H$\beta$, stronger Fe\textsc{ii} sources is naturally interpreted as an increase in $\dot{m}$, consistent with previous observational studies \citep{marziani2003a, zamfir2010, Marziani2018}.

One key feature of the FRADO simulation is that the predicted $\mathrm{FWHM}_{\rm H\beta}$ depends not only on $M_{\bullet}$ and $\dot{m}$, but also on inclination angle. This is a critical aspect in interpreting the observed dispersion in FWHM at fixed $R_{\rm{Fe}}$. At low inclinations (nearly face-on view), the velocity field  is projected less along the line of sight, producing relatively narrow emission lines even for systems with substantial virial velocities. This orientation effect can account for part of the NLS1 population lying at the extreme end of the QMS.

The results support a framework in which a quasar’s location along the QMS is governed primarily by $\dot{m}$ -- as of now a true proxy for $R_{\rm{Fe}}$, then with $M_{\bullet}$ directly responsible for the scatter in FWHM. Eventually $i$ acts as secondary modulator of $R_{\rm{Fe}}$ and FWHM for a fixed $\dot{m}$ and $M_{\bullet}$, respectively. These three major parameters, each of which is discussed in detail below, collectively contribute to shaping the observed structure in the QMS.

\subsection{Physical drivers in the QMS parameter space}

The results indicate that $\dot{m}$ is the principal physical driver of the QMS. In FRADO framework \citep{Czerny2011, naddaf2021, naddaf2022, naddaf2025}, taking $\dot{m}$ as a proxy for $R_{\rm{Fe}}$ naturally reproduces the main observed QMS trend: the transition from PB to PA sources, or, in spectral terms, from weak Fe\textsc{ii}, broad H$\beta$ to strong Fe\textsc{ii}, narrow H$\beta$ \citep{zamfir2010}. This indicates that the QMS is not just empirical but reflects fundamental AGN physics.

\subsubsection{$R_{\rm{Fe}}$ and $\dot{m}$}

The $\dot{m}$ is widely recognized as the primary physical driver of EV1 and the QMS \citep{Boroson2002, Marziani2018, Panda2018}. Quasars with high $R_{\rm{Fe}}$ typically have high accretion rates, often near or above the Eddington limit, affecting both the ionizing SED and BLR microphysics — including temperature, density, and possibly metallicity \citep{Panda2018}. In these high-$\dot{m}$ regimes, radiative forces, particularly dust-driven radiation pressure responsible for H$\beta$ emission, can dominate \citep{naddaf2022, naddaf2025}, driving outflows and blueshifts in both high-ionization lines like C\textsc{iv} \citep{sulentic2007, leighly2018} and, notably, H$\beta$ in the low-ionization regime \citep{naddaf2022, naddaf2025}. The enhanced Fe\textsc{ii} emission in extreme PA (xA) quasars aligns with scenarios where increased radiative cooling is dominant. Such behavior is typical of high-$\dot{m}$ sources, which often exhibit strong outflows and blueshifted emission lines \citep{Marziani2018}. Consequently, $R_{\rm{Fe}}$ serves as an effective observational proxy for $\dot{m}$, reflecting deeper structural changes in the AGN with increasing accretion. The 2.5D FRADO model now provides direct physical proof that $R_{\rm{Fe}}$ alone, without additional parameters, directly traces $\dot{m}$ and vice versa.

\subsubsection{FWHM and $M_{\bullet}$}

The $\mathrm{FWHM}{\rm H\beta}$ has long served as a virial estimator for $M{_\bullet}$, based on the assumption that BLR gas is in virialized motion within the SMBH’s gravitational potential \citep{Peterson2004, vester_peters2006}. Low-ionization lines like H$\beta$ and Mg\textsc{ii} show time delays consistent with a bound region, supporting their origin in a virialized BLR sub-region, as confirmed by reverberation mapping and dynamical modeling \citep[e.g.,][]{Peterson2004, Collin2006, bentz2009}. Their velocity–radius relation, $\Delta V \propto r^{-1/2}$, further indicates virial motion. The presence of this component in both PA and PB sources underscores the role of SMBH dynamics in setting emission-line widths. The xA quasars, which show exceptionally low dispersion in spectral properties, particularly $\mathrm{FWHM}_{\rm H\beta}$, may therefore represent the most virialized and radiatively efficient systems.

\subsubsection{Orientation effect as a geometric parameter}

Orientation introduces an additional layer of complexity in interpreting QMS trends. The FRADO model accounts for the impact of inclination angle by recognizing that the observed line widths and emission strengths can vary significantly depending on the inclination of the accretion disk and associated BLR structures. When viewed at low inclination angles, the projected velocities of BLR clouds are minimized, leading to narrower emission lines. Conversely, high inclinations yield broader lines \citep{wills1986, Shen2014}. This orientation-dependent effect is particularly evident in, but not limited to, the $\mathrm{FWHM}_{\rm H\beta}$ for a given $M_{\bullet}$. It may also affect the $R_{\rm{Fe}}$ for a given $\dot{m}$. The $M_{\bullet}$ and $\dot{m}$ are the primary contributors to the scatter observed along the QMS \citep{marzianietal01,Shen2014}, while $i$, though a secondary factor, acts to broaden this scatter further.

Although orientation is not the main driver of the QMS, it influences spectral appearance and the relative strengths of emission lines, particularly the Fe\textsc{ii}-to-H$\beta$ flux ratio. This arises from differences in the vertical structure and geometry of their line-forming regions within the BLR. Fe\textsc{ii}, dominated by collisional excitation, likely originates in denser, cooler, and flatter regions near the disk plane at larger radii than H$\beta$ \citep{Barth2013, Panda2018, Prince2023}, thus more susceptible to weakening at higher inclinations. H$\beta$, by contrast, likely forms in a more vertically extended, inner region that radiates nearly isotropically and is thus less inclination-sensitive. Consequently, $R_{\rm{Fe}}$ is expected to decline with increasing inclination. These orientation-dependent effects are key to interpreting quasar spectral diversity and separating intrinsic physics from geometry.

We propose the following intuitive conjecture regarding the inclination-dependent modulation for $R_{\rm{Fe}}$. Just as the observed FWHM scales with the $M_{\bullet}$ corresponding intrinsic virial velocity as
\begin{equation}
    \mathrm{FWHM}_{\text{obs}} \propto \mathrm{FWHM}_{\text{int}}(M_{\bullet}) \times f(\sin i)
\end{equation}
where $f$ is a geometry-based function \citep[see e.g.][]{Collin2006} -- in this case equivalent to ``virial factor'', a similar projection effect is also expected to influence the observed $R_{\rm{Fe}}$. Therefore, for a given $\dot{m}$, we expect a relation like
\begin{equation}
R_{\rm{Fe,\,obs}} \propto R_{\rm{Fe,\,int}}(\dot{m}) \times f(\cos i).
\end{equation}
where the dependence on $i$ via $f(\cos i)$ captures the geometric and radiative transfer effects influencing the anisotropic emission of Fe\textsc{ii}. This contributes to a second layer of scatter in the QMS, when the underlying physical parameters are fixed.

This inclination dependence implies that, for a given $\dot{m}$, a highly inclined object will appear left-shifted in the FWHM--$R_{\rm{Fe}}$ plane compared to its position in the FWHM--$\dot{m}$ plane.

\subsection{A concise qualitative look at the role of $Z$}\label{subsec:metallicity}

The FWHM remains nearly unaffected by $Z$ \citep{naddaf2022}, whereas the H$\beta$ flux tends to increase with $Z$ due to larger amounts of recombination-capable gas being lifted and exposed to the ionizing continuum. This causes Fe\textsc{ii} and H$\beta$ fluxes to vary in a broadly concordant manner, keeping $R_{\rm{Fe}}$ approximately constant. The $Z$ also manifests itself through the mean molecular weight ($\mu$) as a function of $Z$ in the definition of the Eddington luminosity \citep{netzer2013book}, and is therefore already implicitly present in $\dot{m}$. 

Although the BLR simulations reported in this work are performed for $5 Z_{\odot}$, in principle, $Z$ in the BLR is expected to evolve through cycles rather than decrease monotonically with redshift \citep{Netzer2007a}, likely reflecting episodic enrichment linked to outflow-driven AGN feedback.
If $Z$, with this potentially episodic behavior, were a dominant driver of $R_{\rm{Fe}}$ alongside $\dot{m}$, it would be expected to introduce a large and relatively uniform inclination-independent scatter in the $R_{\rm{Fe}}$--$\dot{m}$ plane. However, such a strong $Z$-driven scatter is not observed, suggesting that while $Z$ may have a secondary role, it is unlikely to be a primary regulator of $R_{\rm{Fe}}$. 

\section{Conclusion}\label{sec:conclusion}

Our interpretation of $R_{\rm{Fe}} \propto \dot{m}$ provides a physically grounded basis for the empirical EV1 correlations. In the 2.5D FRADO framework, increasing $\dot{m}$ enhances Fe\textsc{ii} emission relative to H$\beta$, consistent with trends in quasar samples. This reinterpretation offers several advantages:
\begin{itemize}
    \item It grounds the QMS in a single physical driver, i.e. $\dot{m}$.
    \item It establishes $\dot{m}$ as the sole physical parameter controlling $R_{\rm{Fe}}$, with increasing $\dot{m}$ leading to a decrease in $\mathrm{FWHM}_{\rm H\beta}$.
    \item It incorporates the $M_{\bullet}$ through the virial $\mathrm{FWHM}_{\rm H\beta}$ .
    \item It reflects the orientation effects, which regulate both FWHM and $R_{\rm{Fe}}$ via line-of-sight dependence.
    \item It links EV1 trends to the underlying physics of BLR.
\end{itemize}

The FRADO model proved to serve as a crucial bridge between empirical observations and the fundamental physical processes driving AGN activity. By correlating spectroscopic features with key physical parameters such as $\dot{m}$, $M_{\bullet}$, and $i$, this model provides an integrated framework that can be used to test theoretical predictions

In the future, by incorporating the Fe\textsc{ii} emission into the FRADO framework, we will quantify the relation between $R_{\rm Fe}$ and $\dot{m}$. We will also conduct a detailed investigation of the role of $Z$, enabling a more complete interpretation of the QMS.

\begin{acknowledgements}
      This project is supported by the University of Liege under Special Funds for Research, IPD-STEMA Program. DH is F.R.S.-FNRS Research Director. BC acknowledges the OPUS-LAP/GA ˇCR-LA bilateral project (2021/43/I/ST9/01352/OPUS 22 and GF23-04053L). MLMA acknowledges financial support from Millenium Nucleus NCN2023${\_}$002 (TITANs), ANID Millennium Science Initiative (AIM23-0001), and the China-Chile Joint Research Fund (CCJRF2310).
\end{acknowledgements}

\bibliographystyle{aa}
\bibliography{naddaf}

\begin{appendix}

\section{Quasar catalogs \& empirical relations} \label{sect:appendixA}

Although we have shown throughout this paper that $R_{\rm{Fe}}$ and $\dot{m}$ are directly correlated in the theoretical context, establishing an empirical relation between them observationally depends sensitively on the chosen sample and the reliability of the underlying measurements. The combination of methodological choices to estimate physical parameters such as $M_{\bullet}$ and $\dot{m}$, data quality issues, and orientation effects can produce significant scatter, potentially distorting the true physical relationship, as we cautioned in Section \ref{sect:rfe_mdot}.
In order to illustrate this, we used the samples from \citet{Hu2008} and \citet{Wu2022}.
\begin{itemize}
    \item The first sample consists of 4,037 quasars at $z < 0.8$ from the 5th Data Release (DR5) of Sloan Digital Sky Survey (SDSS). In their work, the $M_{\bullet}$ is estimated using the continuum luminosity at 5100~\AA\ ($L_{5100}$) and the H$\beta$ line dispersion ($\sigma_{\rm H\beta}$), following the relation proposed by \citet{McGill08}. The $L_{\rm bol}$ is then calculated using $BC=9$. 
    \item The second sample contains 133,018 objects at $z < 1.0$ from the SDSS DR16, of which 86,216 remain after applying the cleaning procedure described by \citet{Wu2022}. In their work, the $M_{\bullet}$ is estimated using fiducial recipes for ``single-epoch virial mass'' as in \citet{vester_peters2006}, which are based on $\mathrm{FWHM}_{\rm H\beta}$ and $L_{5100}$. The $L_{\rm bol}$ is calculated using BCs derived from the mean SED of quasars in \citet{richards2006} for the fiducial  continuum luminosity at rest-frame wavelengths of 5100, 3000, and 1350 $\AA$.
    \item For the \citet{Hu2008} sample, we also independently estimated $M_{\bullet}$ using $L_{5100}$ and $\mathrm{FWHM}_{\rm H\beta}$ following the relation of \citet{bentz2013}, assuming a virial factor of 1 -- as in \citet{dupu2016L}. Hereafter, this estimated set is referred to as ``N+25''. The corresponding $L_{\rm bol}$ values were then obtained using a BC factor of 10. 
\end{itemize}

\begin{figure}
        \centering
        \includegraphics[scale=0.42]{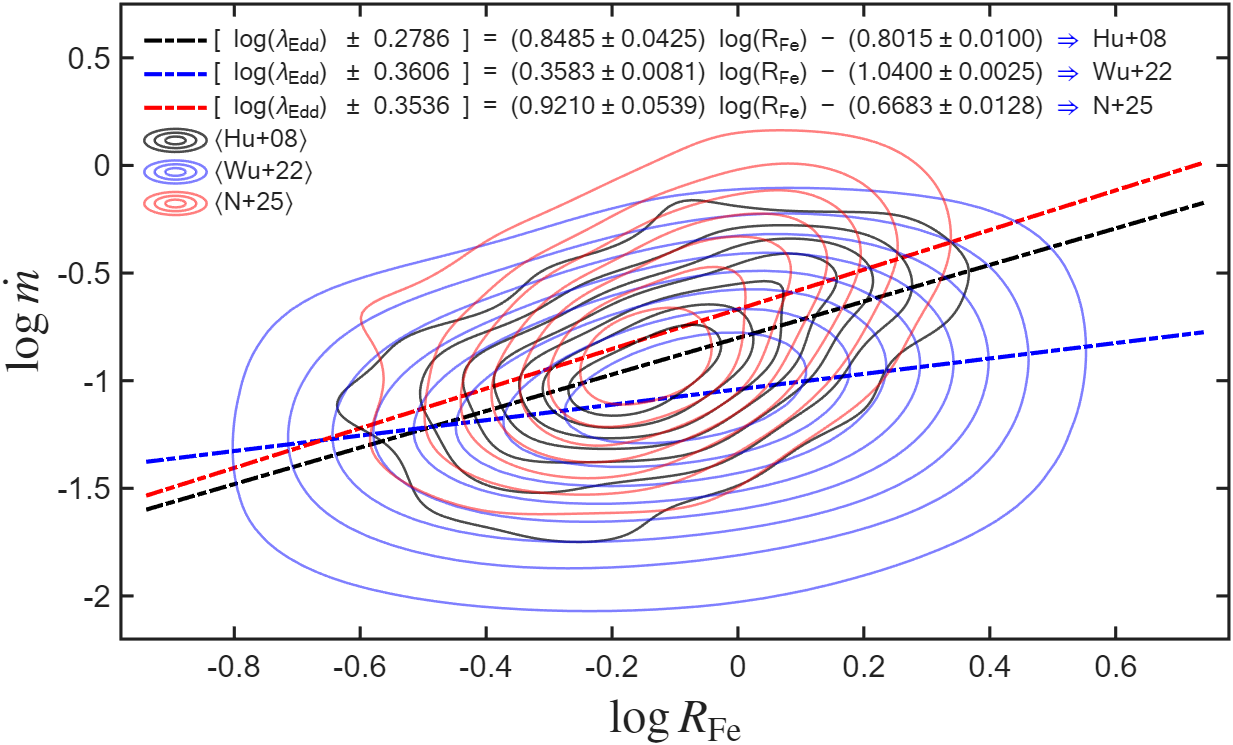}
        \caption{Contour representation of the $\log \lambda_{\rm Edd}$--$\log R_{\rm Fe}$ plane, showing the density distribution of sources for each of the three samples. Black, blue, and red dashed lines are our best-fit linear regressions to datasets from \citet{Hu2008}, \citet{Wu2022}, and N+25, respectively. Slopes and intercepts are shown with their 95\% confidence intervals ($\sim 2\sigma$); numbers in brackets are the RMS scatters of each sample about the respective best-fit lines. Note that $\lambda_{\rm Edd}$ and $\dot{m}$ are used interchangeably; see Appendix~\ref{sec:accretion_definition}.}
    \label{fig:rfe_eddr_contours}
\end{figure}

A comparison of all three datasets is shown in Fig.~\ref{fig:rfe_eddr_contours}, where our best-fit linear regressions are also indicated. The slopes and intercepts of the fits are tightly constrained but show systematic shifts between samples. The scatter of each dataset about its best-fit line, calculated as the root mean square (RMS), ranges from $\sim 0.28$~dex for \citet{Hu2008} to $\sim 0.36$~dex for \citet{Wu2022}. Importantly, the correlation in the sample of \citet{Hu2008} is statistically significant (Spearman's $\rho = 0.451$, $p = 5.1\times10^{-202}$), whereas it is not significant for the \citet{Wu2022} sample ($\rho = 0.258$, $p = 0$).

\begin{figure}
        \centering
        \includegraphics[scale=0.42]{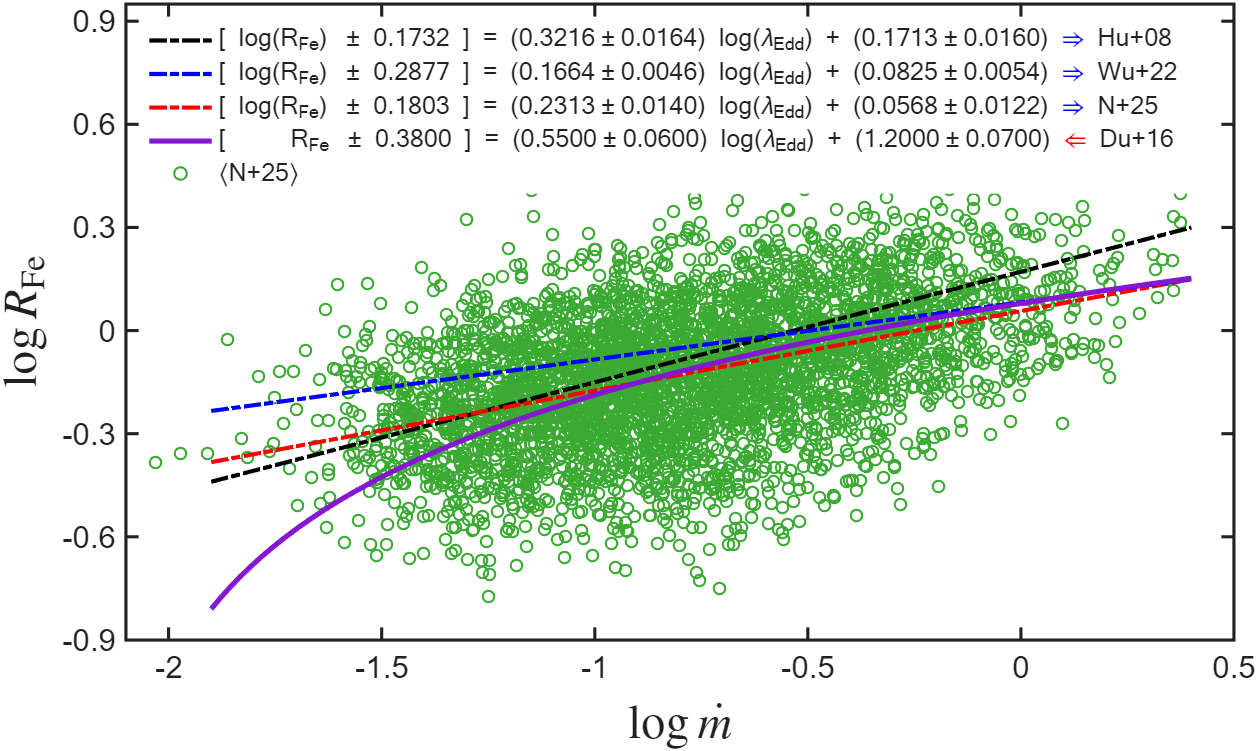}
        \caption{Relations between $\log R_{\rm Fe}$ and $\log \lambda_{\rm Edd}$ for three different samples. Green circles show our N+25 reproduced data . Black, blue, and red dashed lines are our best-fit linear regressions to datasets from \citet{Hu2008}, \citet{Wu2022}, and N+25, respectively. The purple solid curve is the nonlinear prescription from \citet{dupu2016L} found based on a sample of 63 RM super-Eddington quasars. Slopes and intercepts are shown with their 95\% confidence intervals ($\sim 2\sigma$); numbers in brackets are the RMS scatters of each sample about the respective best-fit lines. Note that $\lambda_{\rm Edd}$ and $\dot{m}$ are used interchangeably; see Appendix~\ref{sec:accretion_definition}.}
    \label{fig:rfe_eddr}
\end{figure}

Figure~\ref{fig:rfe_eddr} shows our best-fit linear regressions in the $\log R_{\rm Fe}$--$\log \lambda_{\rm Edd}$ plane for all three datasets. The N+25 data are plotted in the background as a reference. The non-linear relation proposed by \citet{dupu2016L} based on a sample consisting 63 RM Super-Eddington quasars is also over-plotted. Although the correlation in the RM sample of \citet{dupu2016L} is statistically significant (Spearman's $\rho = 0.60$, $p = 2.2\times10^{-7}$) but the sample itself is very small which cannot be practically representative of quasars population.

In all cases, $R_{\rm Fe}$ and $\dot{m}$ are positively correlated; however, the correlation coefficients differ noticeably. The \citet{Hu2008} sample exhibits a steeper trend than \citet{Wu2022}, but shallower than N+25. In contrast, the \citet{dupu2016L} sample displays a distinctly different, curved relationship. This demonstrates that the $R_{\rm Fe}$--$\dot{m}$ correlation depends strongly on the adopted method for estimating the $M_{\bullet}$ and $L_{\rm{bol}}$, thereby, the $\dot{m}$.

We do not compare with the alternative definition of accretion rate, so-called dimensionless accretion rate, $\dot{\mathscr{M}}$, as it involves a simplifying but perhaps misleading convention about the factor of accretion efficiency, $\eta$ (see Appendix~\ref{sec:accretion_definition} for more detail).

\section{Definitions of accretion rate}\label{sec:accretion_definition}

Here we shortly address the alternative definitions around the concept of accretion rate and their corresponding notation and conventions. We specifically focus on dimensionless expressions of this physical parameter, commonly indicated as $\dot{m}$, $\lambda_{\mathrm{Edd}}$, and $\dot{\mathscr{M}}$ in astrophysical notations.
The first two, i.e. $\dot{m}$ and $\lambda_{\mathrm{Edd}}$, are in fact identical, known as ``Eddington ratio'', but expressed in alternative parameter spaces of mass inflow rate and radiative energy outflow rate (luminosity), respectively.
\begin{equation*}
    \dot{m} \equiv \frac{\dot{M}}{\dot{M}_{\mathrm{Edd}}} ,
    \quad \lambda_{\mathrm{Edd}} \equiv \frac{L_{\mathrm{bol}}}{L_{\mathrm{Edd}}},
\end{equation*}
with
\begin{equation*}
    L_{\mathrm{bol}} = \eta \dot{M} c^{2}, 
    \quad L_{\mathrm{Edd}} = \eta \dot{M}_{\mathrm{Edd}} c^{2}.
\end{equation*}

Thus, one can simply show that $\dot{m}$ and $\lambda_{\mathrm{Edd}}$ are identical physical quantities, as in the following 
\begin{equation*}
    \lambda_{\mathrm{Edd}} 
    \equiv \frac{L_{\mathrm{bol}}}{L_{\mathrm{Edd}}} = \frac{\eta \dot{M} c^{2}}{\eta \dot{M}_{\mathrm{Edd}} c^{2}} 
    = \frac{\dot{M}}{\dot{M}_{\mathrm{Edd}}} 
    \equiv \dot{m}.
\end{equation*}

Deviations from this equality occur only when the Eddington mass accretion rate is defined inconsistently. For example, the ``dimensionless accretion rate'', $\dot{\mathscr{M}}$, initially invented to easier capture high accreting super-Eddington sources, use the convention of $\eta = 1$ in the definition of $\dot{M}_{\mathrm{Edd}}$ or alternatively, $L_{\rm bol}$, as below
\begin{equation*}
    \dot{\mathscr{M}} \equiv
    \frac{\dot{M}}{\dot{M}^{\eta=1}_{\mathrm{Edd}}} =
    \frac{\dot{M}}{L_{\mathrm{Edd}}/c^{2}} = \frac{\dot{M} \, c^{2}}{L_{\mathrm{Edd}}} = \frac{L^{\eta=1}_{\mathrm{bol}}}{L_{\mathrm{Edd}}},
\end{equation*}
so the difference between $\dot{\mathscr{M}}$ and Eddington ratio is purely a matter of convention yielding
\begin{equation*}
    \lambda_{\mathrm{Edd}} = \dot{m} = \eta \dot{\mathscr{M}}.
\end{equation*}

This inconsistency leads to inflated estimates and apparent ``super-Eddington'' values that are purely artifacts of the definition. For example, for a typical $\eta = 0.1$, $\dot{\mathscr{M}}$ should be ten times larger than the Eddington ratio, however, it is not as simple as a constant factor of conversion \citep[see][and the references therein for more detail]{wangSEAMBH2014}.

\end{appendix}

\end{document}